\documentclass{sigchi}

\usepackage{balance}       
\usepackage{graphics}      
\usepackage[T1]{fontenc}   
\usepackage{txfonts}
\usepackage{mathptmx}
\usepackage[pdflang={en-US},pdftex]{hyperref}
\usepackage{color}
\usepackage{booktabs}
\usepackage{textcomp}
\usepackage{pdfpages}

\usepackage{todonotes}

\def\plaintitle{Optimizing User Interface Layouts via Gradient Descent}
\def\plainauthor{First Author, Second Author, Third Author, Fourth Author, Fifth Author, Sixth Author}

\def\plainkeywords{Optimization; data-driven design; gradient descent; deep learning; mobile interfaces; LSTM; performance modeling; }

\makeatletter
\def\url@leostyle{%
  \@ifundefined{selectfont}{
    \def\UrlFont{\sf}
  }{
    \def\UrlFont{\small\bf\ttfamily}
  }}
\makeatother
\def\pprw{8.5in}
\def\pprh{11in}

\definecolor{linkColor}{RGB}{6,125,233}
\hypersetup{%
  pdftitle={\plaintitle},
  pdfauthor={\plainauthor},
  pdfkeywords={\plainkeywords},
  pdfdisplaydoctitle=true, 
  bookmarksnumbered,
  pdfstartview={FitH},
  colorlinks,
  citecolor=black,
  filecolor=black,
  linkcolor=black,
  urlcolor=linkColor,
  breaklinks=true,
  hypertexnames=false
}

\begin{document}

\title{\plaintitle}

\numberofauthors{3}
\author{%
  \alignauthor{Peitong Duan\\
    \affaddr{Intel AI}\\
    \affaddr{Santa Clara, USA}\\
    \email{peitong.duan@gmail.com}}\\
  \alignauthor{Casimir Wierzynski\\
    \affaddr{Intel AI}\\
    \affaddr{San Diego, USA}\\
    \email{casimir.wierzynski@intel.com}}\\
  \alignauthor{Lama Nachman\\
    \affaddr{Intel Labs}\\
    \affaddr{Santa Clara, USA}\\
    \email{lama.nachman@intel.com}}\\
}

\maketitle

\begin{abstract}
Automating parts of the user interface (UI) design process has been a longstanding challenge. We present an automated technique for optimizing the layouts of mobile UIs. Our method uses gradient descent on a neural network model of task performance with respect to the model's inputs to make layout modifications that result in improved predicted error rates and task completion times. We start by extending prior work on neural network based performance prediction to 2-dimensional mobile UIs with an expanded interaction space. We then apply our method to two UIs, including one that the model had not been trained on, to discover layout alternatives with significantly improved predicted performance. Finally, we confirm these predictions experimentally, showing improvements up to 9.2 percent in the optimized layouts. This demonstrates the algorithm's efficacy in improving the task performance of a layout, and its ability to generalize and improve layouts of new interfaces.
\end{abstract}




\section{Introduction}
User interface (UI) design is a very difficult process. There are many factors to consider, such as ensuring the UI is efficient to navigate, and that the interface is intuitive so that users can quickly figure out how to use it. A vast number of tools and techniques have been developed to aid designers in this process. They range from evaluative metrics \cite{alemerien2014guievaluator} and models of human performance \cite{li2018predicting,bailly2014model} that designers can use to assess their designs, to tools and techniques that can optimize aspects of the design \cite{todi2016sketchplore, quiroz2007interactive, gajos2004supple, sears1995aide}.
Tremendous progress has been made in modeling human performance on interaction tasks, and recently, deep learning approaches have been introduced, specifically for modelling menu item selection \cite{li2018predicting} and selecting items in a grid-based interface \cite{pfeuffer2018analysis}. These neural network models are able to find complex patterns in large datasets and can be configured to account for various factors affecting task performance time, such as the saliency of the target element for the task, and learning effects from completing a similar task earlier in the sequence. These data-driven models have been shown to outperform analytical models. \cite{li2018predicting} 

Neural networks are differentiable. They are trained to fit a dataset via gradient-based updates to their weights that aim to minimize the difference between the model's predicted value and the observed value in the data. Similar to how a neural network is trained, gradients can also be computed with respect to the network's inputs and be used to update the input to minimize the model's predicted output. This makes neural networks a viable tool for optimization. We decided to apply these neural network models of human performance to the well-studied problem of UI optimization. Specifically, we explore the use of a task performance model's gradients to make updates to user interfaces that aim to minimize an objective function consisting of the model's predicted task completion time and error rate. Since task completion time and error rate are both useful metrics for evaluating UIs, optimizing for them may lead to an interface with better usability \cite{2001}. 

To perform this optimization, we first need a predictive model of task performance. To date, task performance modelling has been done on menus and grid interfaces where tapping and scrolling are the only possible interactions. Thus, we extended the model by Li. et. al. ("Deep Menu") \cite{li2018predicting} to predict task performance times on 2D mobile user interfaces given a UI and a task sequence. Our model also accounts for user error, where the completion time is increased by a penalty if users made a mistake on the task. Hence, our model predicts a metric consisting of both task time and error rate. In addition, our model supports UIs with a variety of element types, such as sliders, icons, and button groups, as well as task sequences with many different interaction types including tapping, dragging and dropping, and sliding (slider bar). Furthermore, our model also handles tasks consisting of multiple interactions. For instance, a task may require the user to tap two different UI elements in sequence. To accommodate this more complex prediction task, we increased the complexity of Deep Menu's architecture and added many more input features, including the location and size of each UI element. 

To scope our work, we focus on tuning the size and location of each element in the UI to minimize task completion time and error rate. We first crowdsourced the completion times  and error rates of a task sequence with 284 tasks on 108 different layout variations of a single user interface, a photo editing UI shown in Figure \ref{fig:figure4}. We then fitted our model to this dataset, achieving an $R^2$ of 0.79 for the target level metric described in \cite{li2018predicting}. 

Then, we developed an optimization algorithm that takes in a user interface and a task sequence and makes iterative adjustments to the x,y position, width, and height of each UI element using gradient descent. We applied this optimization algorithm on various layouts of the photo editing UI. To assess the generalizability of our technique, we applied this algorithm using the same trained model to layouts of a new interface that the model has not been trained on. This is important because designers should not have to collect training data for every UI they plan to optimize. Fortunately, relationships between aspects of the layout and task performance are mostly universal (e.g. the relationship between the UI element's size and the time it takes to point to the element is governed by Fitts' Law \cite{fitts1954information}), so our model should be able to infer these relationships from the dataset and apply it towards optimizing a new interface layout.

This optimization technique produced layouts that have better predicted task performance for both the photo editing UI and the new UI. To verify this experimentally, we crowdsourced task completion times and error rates for a few initial and optimized layouts of both UIs. The observed task performance metric also showed improvements in the optimized layouts, with improvements up to 9.2 percent. These results demonstrate our optimization algorithm's ability to make effective improvements to a layout, as well as its ability to generalize to new interface layouts. From a practical standpoint, our system can facilitate the design process. A designer could start with a set of hand-crafted candidate layouts and use the model to compare their task performance. The designer could also use the optimizer to improve their layouts, generating layout alternatives with better task performance.

To summarize, we made the following contributions:
\begin{itemize}
  \item An extension of the model from Deep Menu to predict task performance of various task types for 2D mobile UI's. We crowdsourced a dataset of task performance times from 379 participants and evaluated our model on this data. 
  \item A new technique for optimizing the layout of mobile UI's using the gradients of a trained task performance predictor network. This technique is generalizable and can improve layouts of a new UI that the model has not been trained on.  
\end{itemize}
\section{Related Work}
\begin{figure*}
  \centering
  \includegraphics[width=1.5\columnwidth]{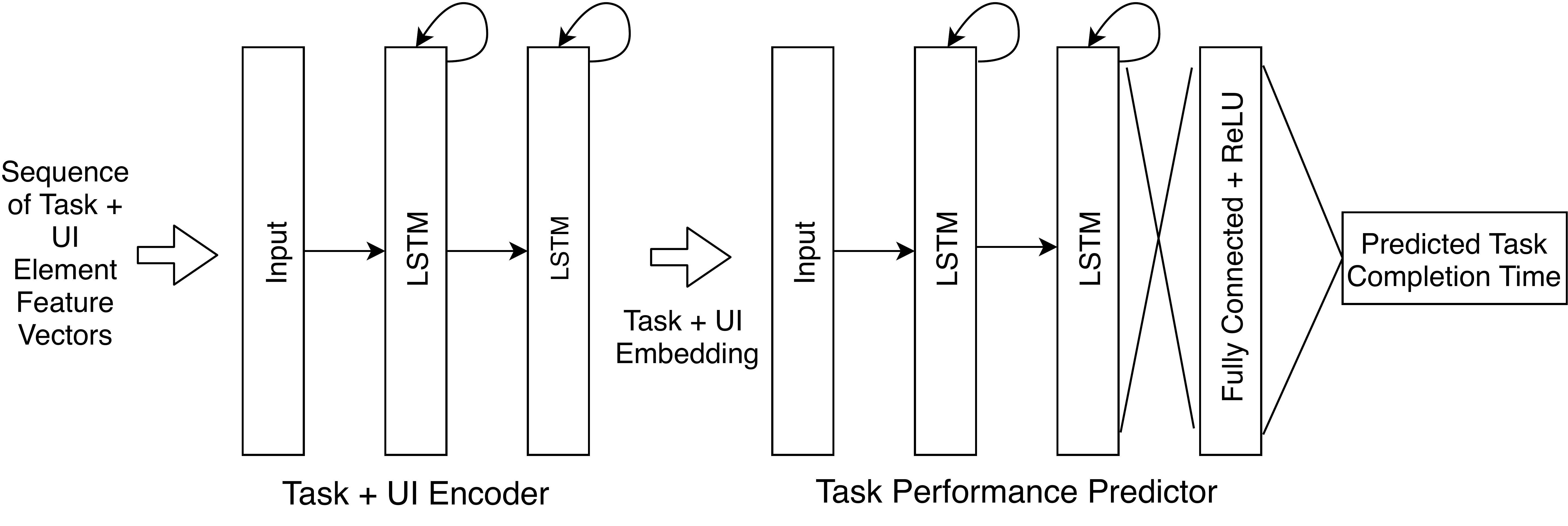}
  \caption{The Task Performance Predictor model's architecture.}~\label{fig:figure3}
  \vspace{-1em}
\end{figure*}
Substantial work has been done in both predicting human performance on interaction tasks and optimizing user interfaces.
\subsection{Modeling Human Behavior}
Modeling human behavior started with simple analytical models that focused on very specific aspects of human performance in isolation. For instance, the well-known Fitts' Law predicts how long it takes users to point to a visual target as a function of distance to the target and target width \cite{fitts1954information}. However, this does not include other factors that affect task performance, such as learning effects from past tasks and visual search time for the target, which have been accounted for by a model on menu item selection proposed by Bailly et. al. \cite{bailly2014model}.

Recently, there has been a shift towards using neural networks to model human behavior, as deep learning models can discover complex patterns in data and do not require extensive feature engineering that is often necessary for analytical models.  In particular, Li. et. al. \cite{li2018predicting} and Swearngin et. al. \cite{swearngin2019modeling} collected large datasets via crowdsourcing on Amazon Mechanical Turk, and then fitted neural network models to their datasets. Li. et. al.'s model (Deep Menu) takes in a menu and a sequence of menu items to select, and predicts the time for each selection. Their model incorporates factors that affect visual search and utilizes recurrent layers to capture learning effects. Swearngin et. al.'s model takes in a mobile UI and an element in the UI and predicts whether users would view the element as tappable. Pfeuffer et. al. also collected a dataset via a 20-user study and fitted a neural network model to predict the time it takes to tap on items in a scrollable mobile grid layout \cite{pfeuffer2018analysis}.
To our knowledge, no one has used deep learning to predict task performance for a general mobile layout, as well as for interaction types beyond tapping and scrolling. Our model expanded the range of interactions to include drag and drop and sliding. We also extended task modeling to incorporate tasks with multiple interactions.
\subsection{UI Optimization}
Because UI design is complex and multifaceted, many techniques have emerged to optimize UI designs, and several different metrics have been used as the objective function. For online games, the goal is usually to maximize user engagement, or how long users spend playing the game. Bayesian optimization \cite{khajah2016designing} and multi-arm bandits \cite{lomas2016interface} have been used to tune features of the game, including font-size and how users enter in input, to maximize user engagement.
For general user interfaces, Krzysztof et. al. created a system (SUPPLE) that automatically generates interfaces, optimizing for a complex function estimating the user effort for a given user trace \cite{gajos2004supple}. For optimizing the UI layout specifically, Quiroz, et. al. used genetic algorithms to evolve the layout of a single UI \cite{quiroz2007interactive}. The color and location of each element in the interface can be evolved, but the layout was restricted to a grid, and changes to the layout consisted of swapping locations of elements within the grid. Furthermore, human input was required to compute the fitness of the layouts in each generation. 

A common technique is to optimize for an objective function combining many layout metrics (e.g. visual clutter). This technique was used in the studies Sketchplore \cite{todi2016sketchplore} and AIDE \cite{sears1995aide}. Sketchplore is perhaps most similar to our work, in that it optimizes an objective function accounting for both usability and aesthetics. Sketchplore's usability component consists of a weighted summation of an analytical model for visual search and Fitts' Law for target acquisition. Our data-driven model expanded upon this by including additional factors that affect task performance. For instance, like Deep Menu, our model also takes in the semantics and saliency of the text labels on each UI element, which should have a strong effect on visual search \cite{nasanen2001effect}, whereas Sketchplore does not factor in text labels in their visual search component. In addition, our task performance predictor also accounts for errors users may make and differentiates amongst types of user interactions, such as drag and drop and sliding, whereas Sketchplore's usability model groups all user interactions as visual search and target-acquisition. 

Until now, no work has been done to apply gradients of a deep neural network model that predicts human task performance to tune the layout of a user interface for better task performance. Li. et. al. did compute gradients of task completion time with respect to input features, but it was to study the model's memory effects. Our work is also the first to use deep learning to predict task completion time and error rate of a general 2D user interface with many different element and interaction types, and also handles multi-step tasks.
\section{Task Performance Prediction Model}
We first expanded the Deep Menu model to predict task performance for mobile user interfaces. Like Deep Menu, the mobile task performance predictor takes in a UI and a sequence of tasks performed on the UI and predicts the completion time of each task. We also increase the task completion time with an error penalty, which is described later. Our model accounts for interfaces with a wide selection of individual and grouped element types. Grouped element types (e.g. a group of icons) consist of a set of elements of the same type that are arranged in a rectangular container (see Figure \ref{fig:figure4} Section E). Individual element types are single elements, such as a slider bar (see Section C). Our model also handles different interaction types, namely tapping, drag and drop, and sliding. We draw largely from Deep Menu's model architecture, and most of our modifications are in the input features.
\subsection{Model Architecture}
Like Deep Menu, we utilize LSTM's capabilities of learning and remembering information from the input task sequence. We have an encoder network that generates an embedding for each task; the embedding is then input into a predictor network, which outputs a prediction for the task's performance time. The network's hierarchical architecture is depicted in Figure \ref{fig:figure3}.
Since the UI may change from user's interactions, features of the UI are fed into the encoder (along with the task information) for every task embedding. In particular, each element in the UI is represented by a fixed length vector, which also contains information about the task. A sequence of these feature vectors, one for each UI element, is input into the encoder network to generate the task embedding. The feature vectors are ordered by the location of the top left corner of each element in a top-down, left-right manner. The input features are discussed in detail in the next section. 

The predictor model takes in a sequence of task embeddings and generates a prediction for the completion time of each task. As shown in Figure \ref{fig:figure3}, the encoder and predictor both have recurrent layers that account for previous tasks in the sequence while predicting the completion time of the current task. This captures the learning effect of users taking less time on tasks as they become more familiar with the UI. Since the tasks and UI for our model are more complex than menu item selection, we have two LSTM layers in both the predictor and encoder models compared to the single recurrent layers in Deep Menu's. The recurrent layers in our predictor model are followed by a feed-forward hidden layer with a ReLU activation function, and the final time prediction is a linear combination of this feed-forward hidden layer; this follows Deep Menu's architecture.
\subsection{Model Features}
For UI element $j$ of task $s$, the feature vector is the following:
\begin{equation}\label{eq:2}
\begin{split}
e_s^j = & [target, len(name), word2vec(name), x, y, width, height,\\ 
& orientation, container\_x, container\_y, container\_width, \\
& container\_height, element\_type]
\end{split}
\raisetag{1\normalbaselineskip}
\end{equation}
These features are selected because they may impact task performance. The first three features ($target$, $len(name)$, and $ word2vec(name)$) are taken from \cite{li2018predicting}, who discussed their effect on task performance. Our $target$ deviates from the definition in \cite{li2018predicting}, and is instead a one-hot vector of length 3 indicating if the UI element is the target for an interaction (e.g. the specified button to be tapped), the drop or sliding target, or not a target. For drag and drop and sliding interactions, one UI element is the target being dragged or slid, and another is the drop target or sliding destination. Hence, these two elements must be differentiated. Likewise, all other categorical features are represented as one-hot vectors. $len(name)$ and $word2vec(name)$ are as defined in \cite{li2018predicting}. $len(name)$ is the length of the text label on the UI element and represents its visual salience. $word2vec(name)$ is the word2vec embedding (reduced to length 4) that captures the semantics of the element's text label. However, in our case, the element may have a symbol or image instead of a text label. In this situation, we would take the $word2vec$ embedding of the word most closely represented by the graphic (e.g. the "undo" embedding is used for the undo icon). $len(name)$ for icons and images is set to a value suitable for it's visual saliency. These features are normalized to a value between -1 and 1. 

An element's location and size affect pointing time \cite{fitts1954information} and visual search \cite{bailly2014model}. Furthermore, the spatial relationships between two related elements affect the performance of tasks requiring interactions with both elements. Hence, we provide the $x$ and $y$ location of the element's (or its bounding box's) center and the element's $width$ and $height$. We also include location and size of the grouped element's container (via $container\_x$, etc.) In addition, for grouped elements, each element in the group has an input vector because individual members are often the task target, as opposed to the group itself. For instance, users would tap one of the buttons in a button group. These spatial features are all in pixels that are then normalized to be between 0 and 1 by the screen width or height. $orientation$ specifies the orientation of the element, which can be vertical, horizontal, or not applicable. Some interactions may be affected by the target element's orientation. For instance, sliding a horizontal slider requires a different motion from sliding a vertical slider and make take a different amount of time to perform. Finally, the element's type (slider, button, icon, etc) is given ($element\_type$) since it determines the element's visual salience, as well as how users would interact with it. 

We then extend our task encoding with the following task-specific features: $[interaction\_type, step, total\_steps]$. $interaction\_type$ indicates the type of interaction required by the task (tapping, drag and drop, etc.), since different interaction types require different gestures (e.g. tapping vs sliding) which should affect completion time. Our model also support tasks consisting of many interactions (e.g. a task may require tapping two different UI elements). These multi-step tasks are presented as a single task to the users, who would have to figure out the individual interactions. Since these multi-step tasks require more cognitive effort from users, they should not be modelled as a series of individual single-interaction tasks, where each step broken down and presented to the user. Our approach to modelling multi-step tasks are as follows: generate a task embedding for each interaction in the task, and use $step$ and $total\_steps$ to identify the embedding as part of a multi-step task. For instance, if a task requires users to first tap element $A$ and then tap element $B$, a task embedding will first be generated for tapping element $A$ with the features $[tap, 1, 2]$ appended, where $[1,2]$ specifies that tapping element $A$ is step 1 of a 2-step task. Similarly the features $[tap, 2, 2]$ are appended to the embedding for tapping element $B$. For single-interaction tasks, $step$ and $total\_steps$ are both set to $1$.

Drag and drop interactions, as well as sliding (the slider handle), involve first acquiring the target and then performing the interaction. Hence, these interactions consist of two steps and are modelled as a 2-step task where the first step is target acquisition and the second is the actual drag and drop or slide. 
\section{Data Collection and Experiments}
In this section, we discuss the dataset we used to evaluate our model, the set-up for crowdsourcing this data on Amazon Mechanical Turk (mTurk), and the results of our data collection. To scope our data collection in a way that is useful for layout optimization, we collected data for various layouts of a single user interface.
\subsection{Dataset}
\begin{figure}
\centering
  \includegraphics[width=0.6\columnwidth]{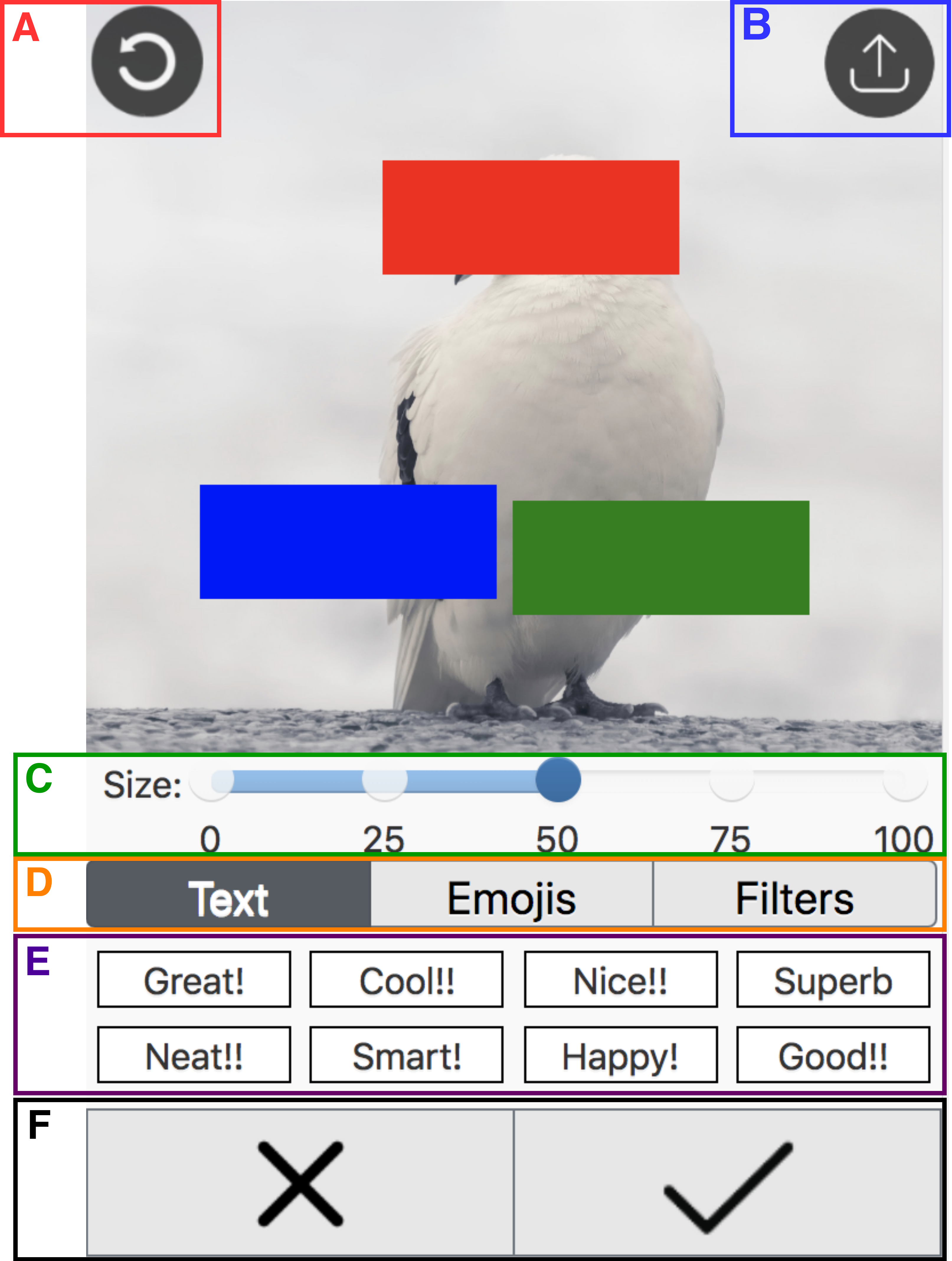}\caption{The photo editing UI with all the UI elements labelled. A is the undo icon, B is the upload icon, C is the slider, D is the button group that controls which set of stickers are displayed, E is the set of stickers (icon group type), and F is the button group with the save (checkmark) and cancel ("X") buttons. The colored rectangles in the photo are not part of the UI; they are the drop targets for Task Type 4 (drag and drop).}~\label{fig:figure4}
    \vspace{-1em}
\end{figure}

\begin{figure*}
  \centering
  \includegraphics[width=1.6\columnwidth]{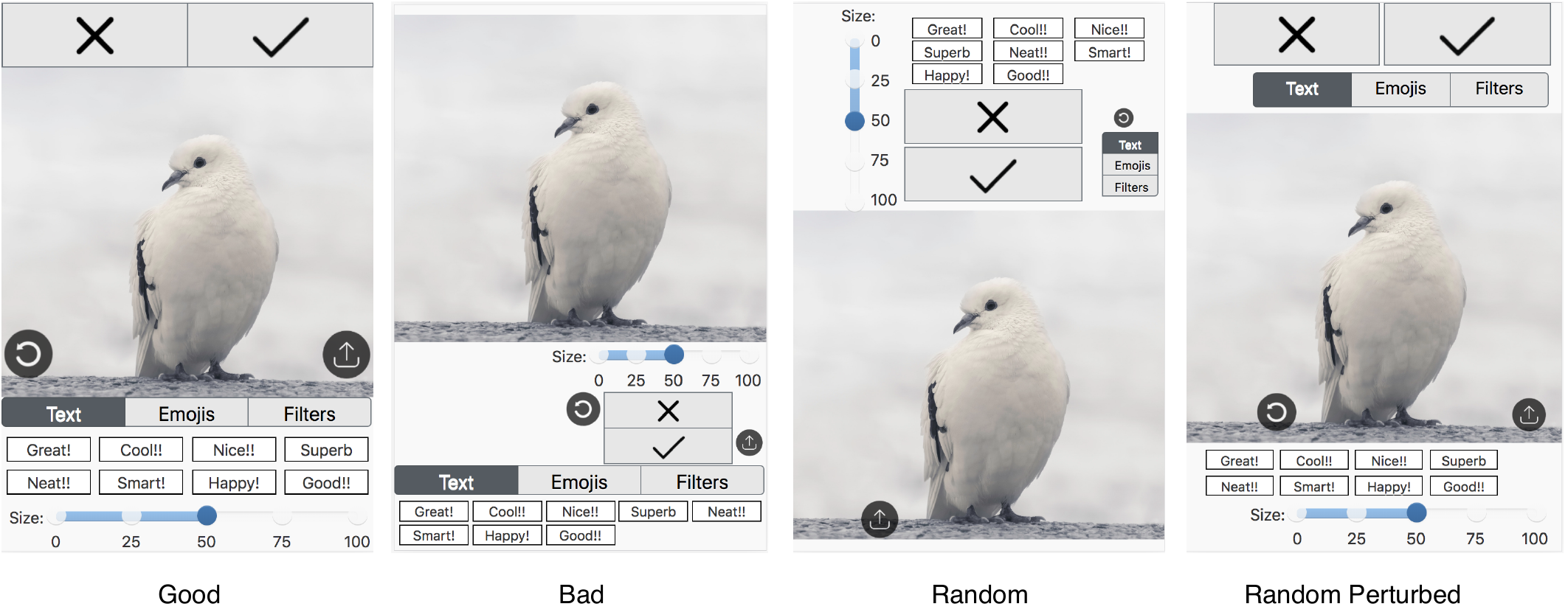}
  \caption{Examples of each layout type in the dataset. The "Random Perturbed" layout is a random perturbation of the "Good" layout. }~\label{fig:figure5}
  \vspace{-1.5em}
\end{figure*}

The dataset consists of 108 different layouts for a user interface, with examples shown in Figure \ref{fig:figure5}. This UI allows users to add stickers and filters to a photo, and consists of the following types of UI elements: button groups (save/cancel buttons and the "Text"/"Emoji"/"Filter" buttons, Sections F and D of Figure \ref{fig:figure4}), icons (undo and upload icons, Sections A and B), icon groups (the stickers, Section E), slider bars (Section C), and static divs (the photo). An icon group is defined as a group of icons confined in a rectangular container with as many icons placed on a row as possible, with uniform and maximal horizontal and vertical spacing between adjacent icons, as shown in Figure \ref{fig:figure4} Section E. Button groups are defined similarly. Icons and buttons are both tappable, but an icon must have a fixed width to height aspect ratio, which is preserved across all 108 layouts. The different layouts in this dataset vary in the size, location, and orientation of each UI element as shown in Figure \ref{fig:figure5}.
\subsubsection{Task Sequence}
Possible user interactions with this photo editing UI include tapping a button or icon, adding a sticker or filter to the photo by tapping on it, adjusting the size of added sticker with the slider, and dragging and dropping the added sticker to a section of the photo. A comprehensive task sequence is created with tasks consisting of individual and combinations of these interactions. Specifically, the different types of tasks include:
\begin{enumerate}
    \item Selecting a sticker or filter to add to the photo
    \item Tapping the appropriate "Text"/"Emoji"/"Filter" button to open the appropriate set of stickers and then selecting the target sticker
    \item Adjusting the size of the added sticker with the slider
    \item Dragging and dropping the added sticker to a specified target in the photo. The drop targets are colored rectangles as shown in Figure \ref{fig:figure4}.
    \item Tapping one of the icons (the undo or upload icon) or one of the buttons (the save or cancel button)
\end{enumerate}
These task types are relatively simple to ensure all mTurk workers perform the same set of interactions in the exact same order as they work through the tasks. This is to guarantee consistent and accurate modeling of each task. We are able to incorporate more complex tasks that require two consecutive interactions (e.g. Task Type 2), which are more difficult to figure out and can be used to assess how intuitive the layout is. For Task Type 4 (drag and drop), the locations of its drop targets are changed (randomly) for every photo to simulate realistic usage of the UI. For Task Type 3, the sliding destination for the slider handle is a specified range on the slider bar (e.g. a value between 50 - 75). Each element in the interface is interacted with at least five times in this task sequence to capture both the learnability of the interface, which is reflected by the first time each task is performed, and the efficiency of the interface, which is shown in later repetitions of the task as the user becomes more familiar with the UI. To simulate realistic usage of this photo editing app, the mTurk participant works through 20 photos in the task sequence and saves or cancels their edits to the current photo before moving onto the next one. This results in a task sequence of 284 tasks.
\subsection{Generating Random Designs}
Out of the 108 different layout of the photo editing UI, 5 are manually designed to meet many of the design guidelines specified by Apple \cite{apple} and Nielsen \cite{nielsen199510} and 3 are manually designed to be bad and violate many of these design guidelines. Figure \ref{fig:figure5} shows an example of a good and bad design. To ensure good coverage of the layout space, the remaining 100 layouts are generated with the layout parameters randomized or by randomly perturbing one of the good designs. 

Fifty designs were generated by randomizing the layout parameters directly. The location, size, and orientation parameters of each UI element are first randomized and the elements are added to the interface one by one, checking for overlap. If the current element being added overlaps with any of the elements already in the UI, its parameters are rerandomized. Otherwise, the element is added to the interface. For grouped elements (e.g. icon groups), the size and location parameters of the container are also randomized. The remaining 50 designs were generated by randomly perturbing each of the 5 good layouts. Since the other set of randomized layouts likely violate design guidelines and are considered bad, only good layouts are perturbed to incorporate more somewhat good layouts to the dataset. For these perturbed layouts, the size of each element is adjusted by a random factor selected uniformly from the interval $[0.7, 1.3]$. Since there is not much white space in these good layouts, the locations are perturbed by randomly swapping adjacent elements with a probability of $0.15$. Figure \ref{fig:figure5} shows examples of these random and perturbed layouts.
\subsection{Data Collection App}
We built a web application using the psiTurk API to crowdsource task completion times and error rates of the task sequence for each layout. psiTurk provides a backend API for recording data and a command line interface to recruit workers from Amazon Mechanical Turk \cite{gureckis2016psiturk}. Following the data collection procedure from \cite{li2018predicting}, the task sequence is presented to the mTurk worker in a the following manner: the worker first sees the instructions for the task, and when users are ready to complete the task, they tap the start button, where they are taken to the UI to complete the task. Our data collection app is described in detail in the supplementary materials, which includes examples of task instructions shown to workers and details on how we handled and recorded the workers' errors.
\subsection{Data Collection Results}
All workers were assigned the same task sequence to work through on one of the 108 different layouts, with at least 3 workers assigned to each. In total, there were 379 participants from Amazon Turk. There were 151 males and 228 females, and around 12 percent of the users were left-handed. In total, we collected completion times of 379 x 284 = 107,636 tasks.

The task completion times for each task were first averaged across all workers for that layout; only completion times where the task was completed correctly were considered. For each task and layout, we remove outliers whose distance is greater than 1.5 median absolute deviations (MAD) from the median. We use MAD because it is more robust in detecting outliers, compared to standard deviation \cite{leys2013detecting}. To incorporate errors, the performance metric assigned to each task and layout is equal to the averaged time for each task per layout increased by an error penalty as shown in the following equation:
\begin{equation}\label{eq:6}
\begin{split}
task\_perf\_metric & = \\
& avg\_task\_time*(1 + 0.5*frac\_err)
\end{split}
\end{equation}
where $frac\_error$ refers to the fraction of workers who made an error on that task out of those who were assigned to that layout. For more severe errors where users incorrectly tapped the save or cancel button, which means they had to redo all the tasks for the current photo, the error penalty was increased from $0.5$ to $0.8$ and $frac\_error$ becomes the fraction of workers who erroneously tapped the save or cancel button. Since these error penalties increase fluctuation in task performance, they decrease the model's prediction accuracy. We carefully tuned these penalty constants to maximize emphasis on errors without sacrificing significant prediction accuracy.

As a sanity check, we computed the average task performance metric for each category of layouts: 563.999 (good, std.err.=10.7), 638.799 (bad, std.err.=46.7), 588.212 (random, std.err.=9.7), and 580.723 (random perturbations of good layouts, std.err.=9.1). As expected, the good designs had on average, a lower value for the task performance metric compared to the bad and random categories, with the bad category having the highest averaged value.
\subsection{Model Performance Results}
This section describes how we configured the model to fit the dataset we collected, and presents the results of an evaluation of the model's accuracy. We also computed the fraction of workers (assigned to each layout) who were left-handed and the average of their ages and appended these statistics to the task embedding. The majority of people use their phones with just their dominant hand and interact with the phone via their thumb \cite{nelavelli2018adaptive}. Since certain elements are located at regions that are more difficult to reach for left-handed users and vice versa, a user's handedness may have an impact on task performance. Furthermore, a person's age is correlated with their UI design preferences, familiarity with mobile technology, and gesture mobility \cite{polyuk_2019}, all of which impact task performance.
\subsubsection{Model Configuration and Loss Function}
There are 8 different element types in the photo editing UI, which means the $element\_type$ feature from Equation \ref{eq:2} is a one-hot vector of size 8. This results in the per-element features vectors defined in Equation \ref{eq:2} to have size 27. The two recurrent layers in the encoder each have 23 LSTM cells. The predictor network has 30 LSTM cells in its two recurrent layers, followed by a feed-forward layer of size 28 to compute the task completion time. To regularize the model and prevent overfitting, we applied a dropout probability of 0.1 to the task embedding and a dropout probability of 0.4 to the feed-forward layer in the predictor (Figure \ref{fig:figure3}). 

We use the same loss function as \cite{li2018predicting}, which is defined as 
\begin{equation}
L_s = \frac{\sum_{i=1}^{|S|} (y_i - t_i)^2}{\sum_{i=1}^{|S|} (y_i - \bar{y})^2}
\end{equation}\label{eq:1}
where $|S|$ refers to the length of the task sequence, $y_i$ is the observed completion time of task $i$, and $t_i$ is the predicted task time. $\bar{y}$ is the average observed task completion time in sequence $S$, so the denominator $\sum_{i=1}^{|S|} (y_i - \bar{y})^2$ is the variance of the task performance times in the sequence. $R^2$ is a standard metric to assess a model's prediction quality and measures the correlation between observed and predicted sequences. This loss function is related to $R^2$ via the equation $R^2 = 1 - L_s$. Since we minimize the loss function $L_s$ during training, we would be maximizing $R^2$.

Our model was implemented in PyTorch, a deep learning framework for Python \cite{paszke2017pytorch}. We trained the network using the Adam optimization algorithm \cite{kingma2014adam} to minimize the loss function with a learning rate of $3\mathrm{e}{-4}$. We also clipped gradients so their norms do not exceed $1.0$.
\subsubsection{Results}
\begin{figure}
\centering
  \includegraphics[width=0.9\columnwidth]{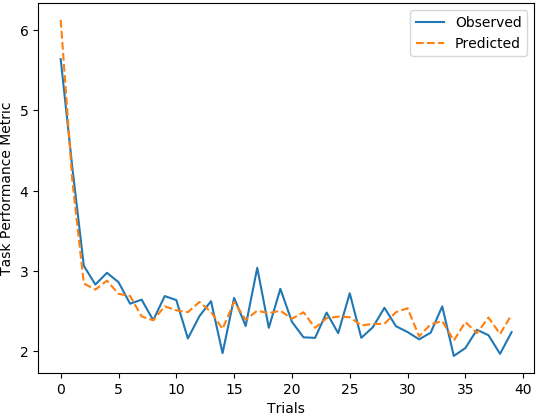}
  \caption{A plot showing the model's accuracy. This graph contains the predicted and observed task performance values for Task Type 3 (sliding), across trials.}~\label{fig:figure7}
  \vspace{-1em}
\end{figure}

We evaluated the model using 6-fold cross validation trained for 850 epochs at each fold, and computed the $R^2$ using the target-level $R^2$ defined by Bailly et. al. \cite{bailly2014model} and used by Li. et. al. to evaluate Deep Menu. This target-level $R^2$ metric examines the relevant task performance for each UI element with varying amounts of practice (trials). Our model achieved a target-level accuracy of 0.79, averaged across all 6 folds. This is comparable to the target-level $R^2$ of 0.76 achieved by Deep Menu on their datasets. Figure \ref{fig:figure7} shows a plot of our model's predicted task performance for Task Type 3 (sliding) and the observed task performance, across trials, demonstrating the model's prediction accuracy.
\section{UI Layout Optimization}
Once we have a trained model to predict task performance, we proceed with layout optimization. We use the gradient descent algorithm, which makes iterative updates to the input based on computed gradients (of the objective function with respect to input) that minimize the objective function. In our case, the input is the location and size of each element in the layout, and our objective function is the model's predicted task performance. This optimization algorithm will be described in detail in the next sections and supplementary materials.

We then used this optimization algorithm to improve layouts of the photo editing UI and another UI that the model has not been trained on, since this technique should be able to generalize and improve layouts of new interfaces to be useful. Designers should not have to collect task performance data for every UI they hope to optimize; the trained model should be able to transfer patterns it learned from one interface to optimize the layout of another. We applied this optimization algorithm on several layouts of both interfaces, which led to predicted improvements in task performance. To verify actual improvements in human performance, we crowdsourced task completion times and error rates for the initial and optimized layouts. The optimized layouts also show improvements in observed task performance for both interfaces.
\subsection{Optimization Algorithm}
Optimization algorithms aim to minimize an objective function, and gradient descent is a particular optimization algorithm that is commonly used to train neural networks. Given objective function $f$ and input $x$, a single update at step $n$ is given by the following equation:
\begin{equation}\label{eq:3}
x_n = x_{n-1} - l_r\nabla f(x_{n-1})
\end{equation}
where $l_r$ is the learning rate that controls the update step size

In our case, the objective function is the sum of the predicted completion times (with error penalty) of all tasks in a task sequence plus the penalty functions, and the inputs are the $x$, $y$, $width$, and $height$ of each element, as well as its corresponding container dimensions. Each layout feature of every element in the UI is updated with their respective gradients individually, following Equation \ref{eq:3}. For each UI element, its features (see Equation \ref{eq:2}) are input into the model for every task in the sequence, which means there will be gradients for the element from each task. Since the size and location of every element in the UI must remain consistent throughout the task sequence, the average is taken over all the element's gradients across tasks, and a single update is made using Equation \ref{eq:3} with this average gradient. The objective function $F$ is formally defined for layout $l$ by the following equation:
\begin{equation}\label{eq:4}
\begin{split}
F(l) & = task\_seq\_perf (l) \\ 
& + (penalty\_constant)(overlap\_penalty(l)) \\
& + (penalty\_constant')(boundary\_penalty(l)) \\ 
& + (penalty\_constant'')(additional\_penalties(l))
\end{split}
\end{equation}
Where $task\_seq\_perf$ is the sum of the predicted task performance of all tasks in the sequence, $overlap\_penalty$ is a differentiable function that is positive if there are overlapping UI elements in layout $l$ and is 0 otherwise. Likewise, $boundary\_penalty$ is a differentiable function that is positive only if any element in $l$ exceeds the boundary of the user interface. $additional\_penalties$ refers to penalty functions of additional constraints the designer hopes to enforce, e.g. ensuring two particular elements are aligned. The penalty constants (e.g. $penalty\_constant$) are constants that add high values to the objective function if their corresponding penalty functions are positive. Furthermore, at each update step, if two elements end up overlapping, their locations are swapped if their gradients indicate that may lower the objective function. 
\subsubsection{Penalty Functions}
\begin{figure*}
  \centering
  \includegraphics[width=2\columnwidth]{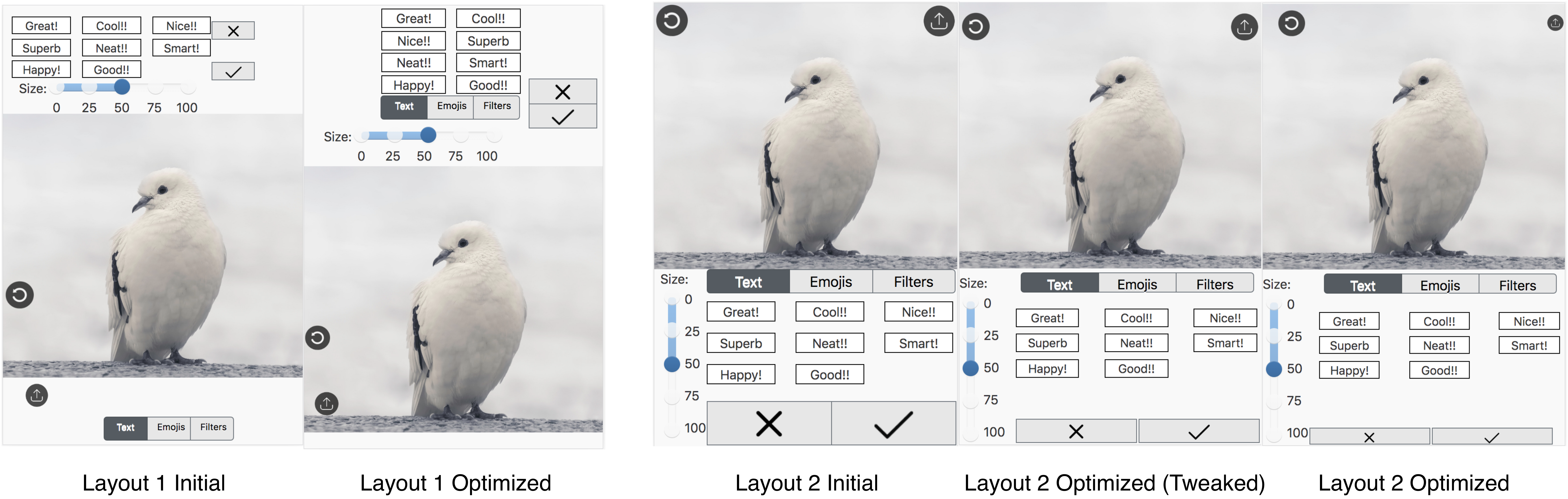}
  \caption{The initial and optimized layouts of two different initial layouts. Layout 1 was optimized with strict constraints, and the predicted task performance improved by 6.3 percent. "Layout 1 Optimized" shows algorithm's output. Layout 2 was optimized using relaxed constraints. The layout directly outputted by the algorithm ("Layout 2 Optimized") had a predicted improvement of 7.1 percent, and the optimized layout after some minor tweaking ("Layout 2 Optimized (Tweaked)") had an improvement of 4.7 percent.}~\label{fig:figure9}
  \vspace{-1.5em}
\end{figure*}
Since each UI element's layout features are updated independently, elements may inevitably overlap with each other or go out the boundary of the interface after updates. Penalty functions can prevent these undesirable conditions by adding a large value to the objective function when these conditions are detected. The penalty function would then steer the gradients towards eliminating these conditions so as to remove the large penalty value. Since penalty functions must contribute to gradients of the objective function to have an effect, they must be differentiable. We use the rectified linear ($ReLU$) function, which is defined as 
\begin{equation}\label{eq:5}
ReLU(x) = max(0, x)
\end{equation}
$ReLU(x)$ is $0$ when $x$ is negative and is equal to $x$ otherwise. The gradient of $ReLU(x)$ is $1$ for positive $x$ and is 0 otherwise. This satisfies the conditions for our penalty function, since it will not contribute to the objective function nor the gradients when it is not activated (equal to 0), and will steer the gradients away from overlaps or other undesirable situations when activated.

The overlap and boundary penalty functions we used are provided, along with derivations, in the supplementary materials. In addition to the necessary overlap and boundary penalty functions, designers can add more constraints to improve the output of the optimization. For instance, penalty functions can be enforced for two UI elements to have the same size, or to introduce a minimum size limit for an element. Furthermore, if a designer wants to group two elements together, such as the sticker button group (Section D of Figure \ref{fig:figure4}) with the stickers (Section E), they can introduce a penalty function to ensure the two elements are in close proximity, aligned horizontally or vertically, and have the same size dimension along their bordering sides (e.g. the sticker button and stickers in Figure \ref{fig:figure4} should have the same $width$s). The equations for all these penalty functions are detailed in the supplementary materials. In sum, penalty functions can be added to ensure any desired characteristics in the optimized layout. The penalty constant ($penalty\_constant$) of each constraint can also be tuned based on importance, with more important constraints being given a larger constant. This is especially useful in the case of conflicting constraints.
\subsubsection{Swapping Locations of UI Elements}
Since the overlap penalty function will push UI elements away from each other if they overlap, elements will be restricted to their initial regions throughout the optimization. For instance, the sticker group (Section E of Figure \ref{fig:figure4}) is constrained to be below the sticker button group (Section D of Figure \ref{fig:figure4}) because whenever the sticker group overlaps with the sticker button group, the sticker group will be pushed back down. To enable exploration of layout alternatives with a much larger location space, we introduce location swapping of overlapping elements. Specifically, when two elements overlap, we look at their gradients of the objective function (Equation \ref{eq:4} ) without the overlap penalty. If the two elements' gradients indicate that swapping their locations may lower the objective function, the swap is performed. A rigorous definition of this swapping condition and details of the location swap can be found in the supplementary materials. 
\begin{figure*}
  \centering
  \includegraphics[width=1.6\columnwidth]{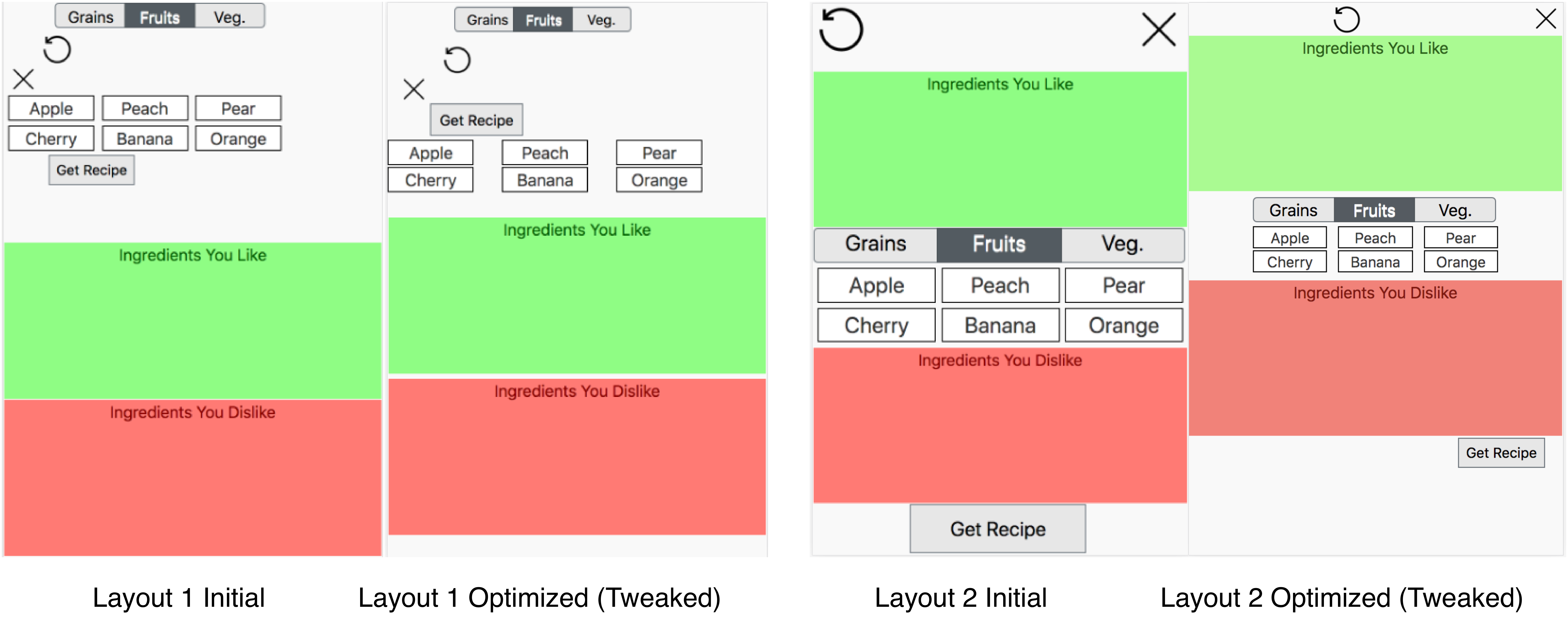}
  \caption{The initial layouts and the optimized layouts after minor tweaking for the recipe planning UI. Layout 1 had a predicted task performance improvement of 3.2 percent, and Layout 2 had a predicted improvement of 5.4 percent. The layouts directly outputted by the algorithm can be found in Figure \ref{fig:figure12}.}~\label{fig:figure11}
  \vspace{-1.5em}
\end{figure*}
\subsection{Optimization Results}
This section contains representative samples of optimization results for the photo editing interface and a new UI that the model has not been trained on. Additional examples of initial and optimized layouts for both interfaces can be found in the supplementary materials. We also crowdsourced human performance data to verify actual improvement in task performance for the optimized layouts.
\subsubsection{Optimization Set-up}
While optimizing the layouts of both user interfaces, we used values of $10000$ for the overlap and boundary penalty constants and smaller values for less serious penalties (e.g. grouping of two elements.) We used a learning rate of $0.05$ for the gradient descent. We also clipped gradients to have norms at most $0.5$ to prevent extreme shifts in the layout from a single update. We ran gradient descent for 500 steps, and saved a CSS file of the layout at each step. We then took the layout with the best predicted task performance. In order to generate layouts with better task performance across the entire population, we passed in the average age of all adult iPhone users (37.7) \cite{statista} and the average fraction of left-handed people in the population (0.1) \cite{hardyck1977left} to our model.
\subsubsection{Photo Editing UI}
For the photo editing UI, we first experimented with optimization using several stringent penalties to ensure good output layouts. These penalties include large minimum size constraints for all UI elements, identical sizes between the undo and upload icons (Sections A and B of Figure \ref{fig:figure4}), and grouping the sticker button with the stickers (Sections D and E of Figure \ref{fig:figure4}). These penalty functions are described in an earlier section "Additional Penalty Functions". Figure \ref{fig:figure9} shows an example of the initial (Layout 1) and optimized (Layout 1 Optimized) layout, along with the predicted task performance improvement. An improvement in task performance refers to a decrease in value of the task completion time with error penalty metric. The improvement in task performance for Layout 1 is likely due to the enlargement of the stickers and other UI elements, which makes tapping easier according to Fitts' Law. Furthermore, the sticker buttons (Figure \ref{fig:figure4} Section D) in the optimized layout are adjacent to the stickers (Figure \ref{fig:figure4} Section E), making the layout more intuitive. This is because sticker buttons control which stickers are displayed, so their relationship is more evident in the optimized layout. Furthermore, this should make tasks where users must first open the appropriate set of stickers before selecting a sticker to add (Task Type 2) more efficient since the two elements are closer together. 

We also applied our technique on a layout that was initially good (Layout 2 of Figure \ref{fig:figure9}). We did not see much improvement when we performed optimization with all the stringent constraints described earlier. Since these strict constraints limit exploration, we removed those constraints and just set low minimum size requirements. With this, we were able to achieve a predicted improvement of 7.1 percent in the total task performance metric for the model output layout (See the "Layout 2 Optimized" in Figure \ref{fig:figure9}). However, some of the elements are too small (e.g. the save and cancel buttons) and misaligned. We can manually enforce the original stringent constraints and align the elements, resulting in a more visually appealing layout (See the "Layout 2 Optimized (Tweaked)"). These alignment and resizing tweaks to the output are very simple and straightforward for the designer to apply, and the tweaked output still had an improvement of 4.7 percent. The slight decline in predicted task performance after tweaking (compared to the algorithm's output layout) could be due to the fact that misaligned elements stand out and are hence, easier to find \cite{balinsky2009aesthetic}. The increased spacing between some elements (e.g. the stickers and exit button (Section F of Figure \ref{fig:figure4})) in the tweaked output layout (compared to the initial layout) likely caused the improvement in predicted task performance.

To verify that our optimized layouts have better human performance, we crowdsourced task completion times and error rates of the initial and optimized layouts for both layouts in Figure \ref{fig:figure9}. For Layout 2, we collected data for the "Optimized Tweaked" layout instead of the "Optimized" layout. We assigned at least 10 mTurk workers to each layout. The crowdsourced results confirm improvement in task performance in the optimized layouts. The observed performance increase (decrease in the value of the task performance metric) is 8.9 percent for Layout 1 (6.3 percent predicted) and 2.0 percent for Layout 2 (4.7 percent predicted). Layout 2 likely had a smaller improvement than Layout 1 because Layout 2 was initially quite good. To verify statistical significance of the observed performance improvement, we computed a two-sample t-test for the observed task performance metric for the initial and optimized versions of both layouts. For Layout 1, $t(13) = -16.674$, $p < 1\mathrm{e}{-13}$ and for Layout 2, $t(10) = -2.780$, $p < 0.05$, which indicate statistical significance of the observed improvements. Hence, the crowdsourced results confirm task performance improvements in the optimized layouts.
\subsubsection{New UI: Recipe Planner}
\begin{figure}
  \centering
  \includegraphics[width=0.75\columnwidth]{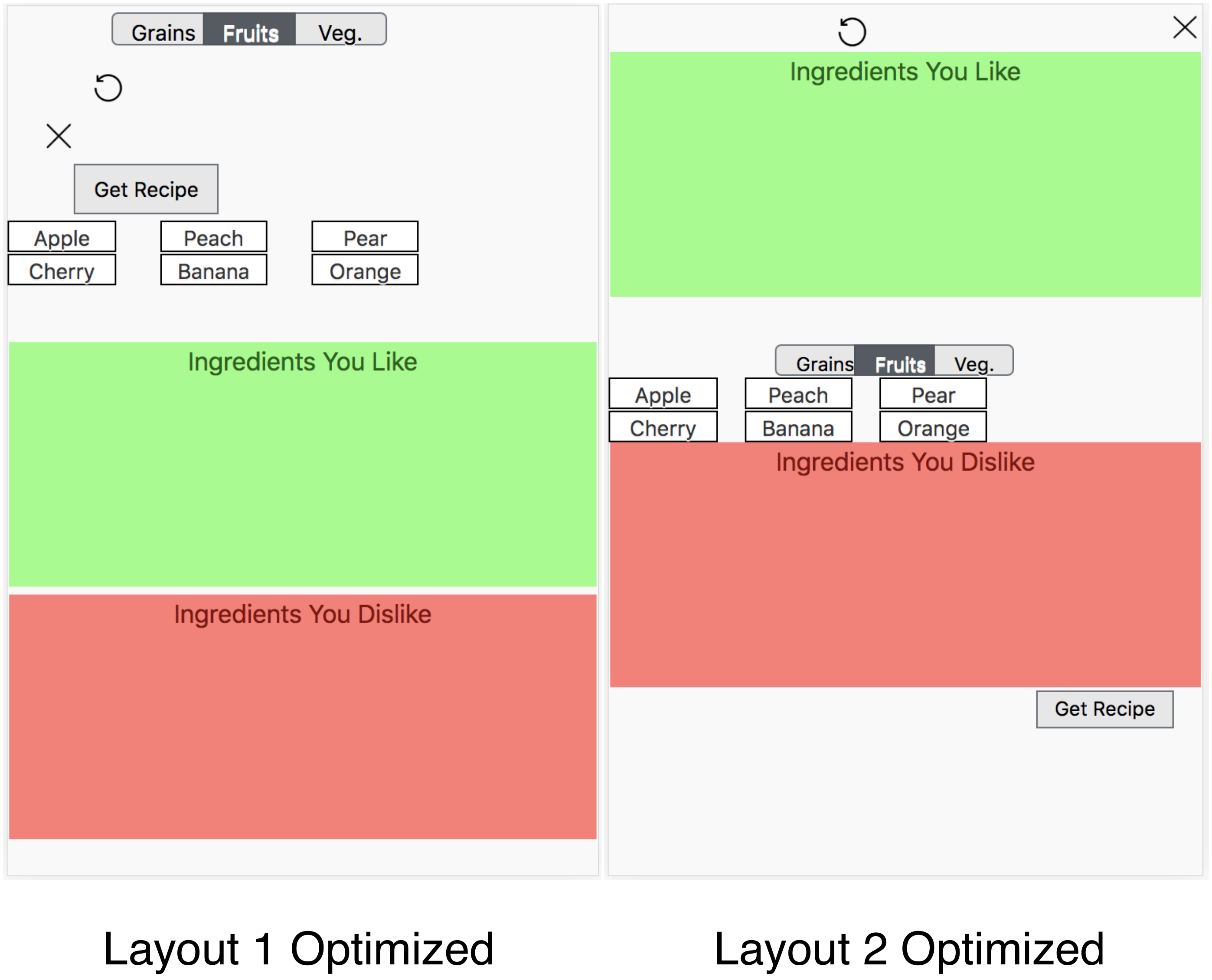}
  \caption{Layouts directly outputted by the algorithm. For Layout 1, the algorithm made a 5.8 percent improvement (predicted) in task performance, and for Layout 2, the algorithm made a 4.1 percent improvement. These output layouts were tweaked slightly to for alignment and re-sizing. Figure \ref{fig:figure11} shows the tweaked layouts.}~\label{fig:figure12}
  \vspace{-1.5em}
\end{figure}

The recipe planner is an interface where users drag and drop ingredients that they like and dislike to the appropriate boxes (see Figures \ref{fig:figure11} and \ref{fig:figure12}). Once users have made all their selections, they tap the "Get Recipe" button to get a list of recipes they may like based on their ingredient preferences. All element and interaction types in this recipe planning UI are found in the photo editing UI. This is necessary in order to utilize the model trained on the photo editing UI to optimize layouts of the recipe planning UI, as the model would probably not be able to make accurate predictions on new element nor interaction types. However, the UI element and interaction types in the photo editing UI are very general and can be used to build a wide variety of user interfaces. 
We created a comprehensive task sequence for the recipe planning interface with which we will optimize for task performance. The task types include tapping the "Get Recipe", undo, cancel, or one of the ingredients buttons ("Grains", "Fruits", "Veg."), dragging and dropping a specified ingredient sticker to the "Ingredients You Like" or "Ingredients You Dislike" box, and a two-step task, where the user must first open the appropriate set of ingredients stickers via tapping the corresponding ingredients button and then dragging and dropping the target ingredient to the appropriate box. This task sequence for the recipe planner UI had 136 tasks.

We first tried optimizing the layout with strict constraints, such as large minimum size constraints for the UI elements. However, these strict constraints did not lead to an improvement in predicted performance of the task sequence. We then relaxed these constraints to include only low minimum size constraints. This resulted in improvements in predicted task performance of the task sequence, even after the optimization output has been tweaked to align elements and enlarge elements that were too small. The optimization results of an initially bad layout (Layout 1) is shown are Figures \ref{fig:figure11} and \ref{fig:figure12}. For Layout 1, the optimization algorithm moved the ingredients stickers closer to the "Ingredients You Like" and "Ingredients You Dislike" drop targets and added more spacing between the undo and cancel icons in Layout 1 that were initially too close together.
We also optimized a layout that was initially good (Layout 2), and Figures \ref{fig:figure11} and \ref{fig:figure12} show the results. For Layout 2, the algorithm decreased the distance users had to drag the ingredient stickers to the drop targets, and also moved the "Get Recipe" button and the undo icon farther to the right, which makes it more accessible to right-handed users, who make up the 90 percent of the user base.
To verify that the optimized layouts have improved human task performance, we crowdsourced task performance times and error rates for Layouts 1 and 2. There is an observed improvement of 9.2 percent ($t(12) = -9.799, p < 1\mathrm{e}{-8}$) and a 3.2 percent predicted improvement for Layout 1 and an observed improvement of 4.9 percent ($t(13) = -5.622, p < 1\mathrm{e}{-5}$) and a 5.4 percent predicted improvement for Layout 2.
\section{Discussion}
Our optimization algorithm was able to discover layout alternatives with  better task performance for both user interfaces, which demonstrates the algorithm's generalizability. Hence, it would likely be able to improve the layout of any interface containing UI elements and interactions found in the photo editing interface, which is a very large set of possible UI's. Furthermore, if the task performance predictor is trained on a large enough dataset with examples of all UI element types and interaction types, our optimization algorithm may be able to make improvements to the layouts of any general mobile UI.

The findings of our optimization experiments support those by Lomas et. al. \cite{lomas2016interface}. Their paper emphasized optimizing for the right metric, and while our optimization technique returned layouts with better task performance, the layouts did not show much improvement in terms of aesthetics. Hence, our objective function could have been more comprehensive and incorporated metrics for the visual aspect of the layout. Furthermore, as shown in the layouts directly outputted by the optimization algorithm (Figures \ref{fig:figure11} and \ref{fig:figure12}), the model made some elements very small to possibly reduce user error. Lomas et. al. warned that the optimization output may be unsatisfactory when metrics are optimized to the extreme.
\subsection{Human-AI Collaboration}
These limitations with our optimization algorithm can be fixed easily by humans, as shown in Figures \ref{fig:figure11} and \ref{fig:figure12}. Both human judgement and the optimization algorithm can be misleading, so ideally humans and AI would collaborate in the design process \cite{lomas2016interface}. The following illustrates how this hybrid workflow might work with our system. Designers can start by building an initial layout and specifying any constraints in the optimization output, such as elements that must be grouped together. Then, our algorithm returns a layout that the designer can refine for aesthetic quality and other features. The designer can then use the task performance predictor to compare the task performance between the refined and initial layouts. Furthermore, the designer can run the optimization again on the refined layout to further improve its task performance, and this iteration between AI optimization and human refinement can be continued until a satisfactory layout has been achieved. Furthermore, the optimization algorithm outputs the CSS file for the layout at every step, which generates a large set of layout alternatives. If the designer sees a layout they like better than the one with the best predicted task performance, they can work with their preferred layout instead. In sum, there are many ways for our optimization algorithm to assist designers in creating a great layout for their user interface.
\subsection{Limitations and Future Work}
As mentioned earlier, one limitation is that our optimization technique does not account for layout aesthetics. Less effort will be required from designers if the algorithm's output had better aesthetics. Another limitation is that many of the tasks in the dataset's task sequence are simple and require only one interaction. Although we did have 2-interaction tasks, including even more complex tasks could result in stronger emphasis of the layout's learnability in the task performance. 

The limitations suggest promising opportunities for future work. There exist metrics for aesthetic quality, such as the Balinsky symmetry metric that measures the grid quality of the layout \cite{balinsky2006evaluating}, which can be incorporated in the objective function to optimize for the appearance of the layout in addition to usability. Furthermore, since our system provides a platform for human-AI collaboration in UI design, it would be insightful to study the dynamics between human and AI during such a collaboration, such as disagreements that may arise between the two. It will also be interesting to compare the final layout after many iterations of this human-AI collaboration for a novice designer with the layout created by an expert designer.
\section{Conclusion}
Our model predicts task performance (a metric combining task completion time and error rate) of a general mobile UI that contains a variety of element types and supports many types of user interactions. We then developed an algorithm that uses this model to make iterative updates to a UI's layout using gradient descent. We used this optimization algorithm to improve layouts of two UI's, including one interface that our model has not been trained on. Our optimization algorithm was able to generate layout alternatives with better task performance for both interfaces, as confirmed by crowdsourcing studies comparing the initial and optimized layouts. This demonstrates the effectiveness of this algorithm in improving a layout's task performance, as well as its ability to generalize and make improvements to new interface layouts. The optimization's output layout do need minor human refinement in order to look aesthetically pleasing. However, collaboration between human and AI is recommended over relying solely on either. 
\balance{}

\bibliographystyle{SIGCHI-Reference-Format}
\bibliography{proceedings}

\end{document}